\documentclass[iop]{emulateapj}      

\usepackage{graphics} 
\usepackage{amsmath} 

\newcommand{\um}{${\rm \mu m}$~}
\newcommand{\mm}{${\rm \mu m}$} 

\long\def\symbolfootnote[#1]#2{\begingroup%
\def\thefootnote{\fnsymbol{footnote}}\footnote[#1]{#2}\endgroup}

\slugcomment{AJ accepted.}

\shorttitle{ALMA 1.3 mm Map of the HD~95086 System}
\shortauthors{Su et al.}

\begin{document}

\title{ALMA 1.3 Millimeter Map of the HD~95086 System}

\author{Kate Y. L. Su\altaffilmark{1,2}, Meredith A. Macgregor\altaffilmark{3,14}, Mark Booth\altaffilmark{4}, 
David J. Wilner\altaffilmark{3}, Kevin Flaherty\altaffilmark{5}, A. Meredith Hughes\altaffilmark{5}, 
Neil M. Phillips\altaffilmark{6}, Renu Malhotra\altaffilmark{7}, Antonio S. Hales$^{8,9}$, 
Sarah Morrison\altaffilmark{7,15}, Steve Ertel\altaffilmark{1}, Brenda C. Matthews\altaffilmark{10,11}, William R. F. Dent$^{8}$, 
Simon Casassus$^{12,13}$}
 
\affil{$^1$ Steward Observatory, University of Arizona, 933 N Cherry Ave., Tucson, AZ 85721, USA}
\affil{$^2$ Institute of Astronomy and Astrophysics, Academia Sinica, P.O. Box 23-141, Taipei 106, Taiwan}

\affil{$^3$ Harvard-Smithsonian Center for Astrophysics, 60 Garden Street, Cambridge, MA 02138, USA}
\affil{$^4$ Astrophysikalisches Institut und Universit\"atssternwarte, Friedrich-Schiller-Universit\"at Jena, Schillerg\"a{\ss}chen~2--3, 07745 Jena, Germany}              
\affil{$^5$ Department of Astronomy, Van Vleck Observatory, Wesleyan University, Middletown, CT 06459, USA}
\affil{$^6$ Joint ALMA Observatory (JAO); European Southern Observatory}
\affil{$^7$ Lunar and Planetary Laboratory, University of Arizona, Tucson, AZ 85721, USA}
\affil{$^{8}$ Joint ALMA Observatory, Alonso de C\'ordova 3107, Vitacura 763-0355, Santiago, Chile}
\affil{$^{9}$ National Radio Astronomy Observatory, 520 Edgemont Road, Charlottesville, Virginia, 22903-2475, USA }

\affil{$^{10}$ National Research Council of Canada Herzberg Astronomy and Astrophysics Programs, Victoria, BC, V9E 2E7, Canada}
\affil{$^{11}$ University of Victoria, 3800 Finnerty Road, Victoria, BC, V8P 5C2, Canada}

\affil{$^{12}$ Departamento de Astronomia, Universidad de Chile, Casilla 36-D, Santiago, Chile }
\affil{$^{13}$ Millennium Nucleus ``Protoplanetary Disks'', Santiago, Chile }
\affil{$^{14}$ Current address: Department of Terrestrial Magnetism, Carnegie Institution for Science, 5241 Broad Branch Road, Washington, DC 20015, USA}
\affil{$^{15}$ Current address: Center for Exoplanets \& Habitable Worlds, Pennsylvania State University, University Park, PA 16802, USA }

\begin{abstract}

Planets and minor bodies such as asteroids, Kuiper-belt objects and
comets are integral components of a planetary system. Interactions
among them leave clues about the formation process of a planetary
system. The signature of such interactions is most prominent through
observations of its debris disk at millimeter wavelengths where
emission is dominated by the population of large grains that stay
close to their parent bodies. Here we present ALMA 1.3 mm observations
of HD~95086, a young early-type star that hosts a directly imaged
giant planet b and a massive debris disk with both asteroid- and
Kuiper-belt analogs.  The location of the Kuiper-belt analog is
resolved for the first time. The system can be depicted as a broad
($\Delta R/R \sim$0.84), inclined (30\arcdeg$\pm$3\arcdeg) ring with
millimeter emission peaked at 200$\pm$6 au from the star. The 1.3 mm
disk emission is consistent with a broad disk with sharp boundaries
from 106$\pm$6 to 320$\pm$20 au with a surface density distribution
described by a power law with an index of --0.5$\pm$0.2. Our deep ALMA
map also reveals a bright source located near the edge of the ring,
whose brightness at 1.3 mm and potential spectral energy distribution
are consistent with it being a luminous star-forming galaxy at high
redshift. We set constraints on the orbital properties of planet b
assuming co-planarity with the observed disk.

\end{abstract}


\section{Introduction}


Debris disks were discovered by {\it IRAS} \citep{aumann84} as
infrared excess emission from dust orbiting stars and sustained by
collisions of leftover planetesimals and cometary activity. They often
have a structure analogous to that of minor body belts in the solar
system, with asteroid- or Kuiper-belt components. The majority of the
known debris disks are massive, Kuiper-belt analogs not only because
the collisional evolution proceeds more slowly at large orbital
distances but also stars are faint in the far-infrared, making
positive identifications of excess from cold debris much easier. It is
interesting to note that the first Kuiper Belt Objects in our own
solar system were not discovered until 1992 \citep{jewitt93}, eight
years later than the {\it IRAS\ } discovery.

Planets, minor bodies, and leftover planetesimals all form as a
consequence of agglomeration processes that occur within the
protoplanetary disk.  Interactions between them during the formation
and subsequent evolution leave signs in the disk that can be used to
study the current state and past history of a planetary
system. Therefore, these faint dusty disks are excellent tools to
understand the outer zones of exoplanetary systems including our own.

With sensitive infrared surveys, hundreds of debris disks are known
\citep{matthews14}, providing a rich resource to study planetary
system evolution and architecture.  Although thousands of exoplanets
and candidates have been discovered through radial velocity and
transit measurements, this breakthrough is currently biased toward the
inner zones of systems, not sensitive to planets like Jupiter and
Saturn beyond 5 au. Recent improvements in high contrast imaging have
enabled us to find planets out at the same stellocentric distance scales 
as the debris disks. Fomalhaut \citep{kalas08}, HR~8799
\citep{marois08,marois10}, $\beta$ Pic \citep{lagrange09}, HD~95086
\citep{rameau13}, HD~106906 \citep{bailey14}, and 51 Eri
\citep{machintosh15} are prominent examples of such systems known to 
host both debris disks and directly imaged planets.

From the observed dust temperatures derived from disk spectral energy
distributions (SEDs) of $\sim$200 debris disks, \citet{ballering13}
report a weak trend that the inner edge of the cold planetesimal zone
appears to depend on the luminosity/temperature of the star,
indicating a signpost for planetary migration and/or
shepherding. However, disk extents estimated from SEDs are
degenerate. Any inferred radii depend strongly on the assumed
composition and the particle size distribution. Due to this
degeneracy, it is very difficult to translate SED measurements into
physical sizes directly. Even when there are resolved images available
(mostly in the far-infrared), the exact location of the parent bodies
is still uncertain due to the effect of non-gravitational forces
(radiation and drag) on small grains. The true parent-body
distribution in debris disks can be provided by resolved
submillimeter/millimeter images which probe large (mm-size) grains
that stay close to their parent bodies. Disk morphologies suggestive
of influences from unseen planets, such as resonance clumps
\citep{wyatt03} and/or apo-center glow \citep{pan16}, are also best
observed at submillimeter/millimeter wavelengths (e.g.,
\citealt{ertel12,lohne17}).  Existing ALMA data on debris disks show a
large variety of Kuiper-belt analogs: some systems have very narrow
rings of parent bodies (e.g., Fomalhaut,
\citealt{boley12,macgregor17}, and $\epsilon$ Eri, \citealt{booth17}),
and some have either multiple rings (HD~107146, \citealt{ricci15a}) or
broad disks (HR~8799, \citealt{booth16}; $\tau$ Ceti,
\citealt{macgregor16b}; 61 Vir, \citealt{marino17}).  The parent body
distributions are therefore giving insights to the possible overall
structure of the planetary systems.

HD~95086 is a young (17$\pm$4 Myr, \citealt{meshkat13}) A8 star that
possesses a large infrared excess, indicative of a massive debris disk
\citep{chen12}, and a $\sim$5 $M_J$ planet at the projected distance
of $\sim$56 AU \citep{rameau13,rameau16}. 
Compared to the Hipparcos catalog, the Gaia DR1
catalog gives a slightly closer distance, 83.8$\pm$1.9 pc
\citep{gaia16}, which we adopt throughout the paper. Its disk was
marginally resolved by {\it Herschel\ } and found to be inclined at
$\sim$25$^{\circ}$ from face-on \citep{moor13}. Analysis of its
detailed infrared SED and re-analysis
of the resolved images suggest that the debris structure around HD~95086 
is very similar to that of HR~8799: a warm ($\sim$170 K) belt, a
cold ($\sim$60 K) disk, and an extended disk halo (up to $\sim$800 AU)
\citep{su15}. Modeling the disk surface brightness distribution at 70
and 160 $\mu$m suggests that the extended emission seen in the
far-infrared is largely from the small grains produced by frequent
collisions due to dynamical
stirring of planetesimals and launched from the system by stellar
radiation in the form of a disk halo. Therefore, the inclination derived
from the {\it Herschel} images might not be a true representative of 
the planetesimal disk or be subject to a large error.  It is then
crucial to measure the intrinsic distribution of the planetesimal population,
as traced by millimeter emission from large grains, in order to properly
characterize the possible perturbers, HD~95086 b and any unseen planet(s)
interior to the cold disk.

Here we present the first millimeter observations of the HD~95086
system, obtained by the Atacama Large Millimeter/submillimeter Array
(ALMA). Our observations reveal the location of the cold Kuiper-belt
analog for the first time. The paper is organized as follows. Details
about the observations and general data reduction are given in Section
2. In Section 3, we first present the dust continuum map of the system
which can be described as an inclined ring plus a bright point source
near the outer edge of the ring. We then determine the properties of
the disk (flux and geometry) and those of the bright source (flux and
position) using both visibilities and imaging model approaches. In
Section 4, we revise the disk SED based on the new
observations, discuss the ring's width and possible asymmetry, the
likely nature of the bright source, and obtain new constraints on 
HD~95086 b. Conclusions are given in Section 5.

\begin{deluxetable*}{lccccrccc}
\tablewidth{0pc}
\footnotesize 
\tablecaption{Observational Log \label{tab1_obs}}
\tablehead{
\colhead{Date} & \colhead{Block UID}  & \colhead{\# of Used} & \colhead{Baselines} & \colhead{PWV}& \colhead{T$_{sys}$} & \colhead{Time on Source}  & \colhead{Hour Angle} & \colhead{Flux Calibrator}\\ 
\colhead{}   & \colhead{}  & \colhead{Antennae}   & \colhead{[m] }   & \colhead{[mm]} & \colhead{[K]} & \colhead{[min]} & \colhead{at mid-point} & \colhead{} 
}
\startdata 
\multicolumn{3}{l}{\underline{\#2013.1.00773.S, data set A}}  &    &   &  &   &  & \\ 

2015-01-28  &  X1beb   &  38  & 15.1 -- 348.5   &    1.27  &  83.7  &  45.86  & +02:25  & J1107 \\
2015-04-04  &  Xba2    &  39  & 15.1 -- 327.8   &    1.02  &  73.8  &  45.86  & --00:38 & Callisto \\
2015-04-05  &  X267e   &  39  & 15.1 -- 327.8   &    1.28  &  77.2  &  45.86  & --00:09 & Ganymede \\
2015-04-05  &  X2a9e   &  39  & 15.1 -- 327.8   &    1.27  &  77.3 &  45.86  & +01:27  & J1107 \\
2015-04-05  &  X2e6d   &  39  & 15.1 -- 327.8   &    1.25  &  80.7  &  45.86  & +02:55  & Titan \\
2015-04-06  &  X14f2   &  36  & 15.1 -- 327.8   &    1.23  &  76.5  &  45.86  & +01:47  & J1107 \\
\\
\multicolumn{3}{l}{\underline{\#2013.1.00612.S, data set B}}  &    &   &  &   &  & \\ 
2015-04-10  &  X1412   &  35  & 15.3 -- 348.5   &    2.14  & 101.0  &  45.36  & +00:14  & Ganymede \\
2015-04-10  &  X1d34   &  35  & 15.3 -- 348.5   &    2.38  & 108.9  &  45.36  & +02:16  & J1107 \\
2015-04-14  &  Xbb6    &  36  & 15.3 -- 348.5   &    3.65  & 144.6  &  45.36  & --01:18 & Ganymede \\
2015-04-23  &  X1462   &  39  & 15.1 -- 348.5   &    1.88  & 103.3  &  45.36  & +03:12  & Titan \\
2015-05-01  &  X883    &  37  & 15.1 -- 348.5   &    1.95  & 100.7  &  45.36  & +00:55  & Ganymede \\
2015-05-02  &  Xd15    &  37  & 15.1 -- 348.5   &    1.18  &  85.0   &  45.36  & +00:15  & Ganymede\\
\enddata 
\end{deluxetable*}  

\begin{figure*}[thb] 
  \figurenum{1}
  \epsscale{1.15}
  \label{alma_images} 
  \plotone{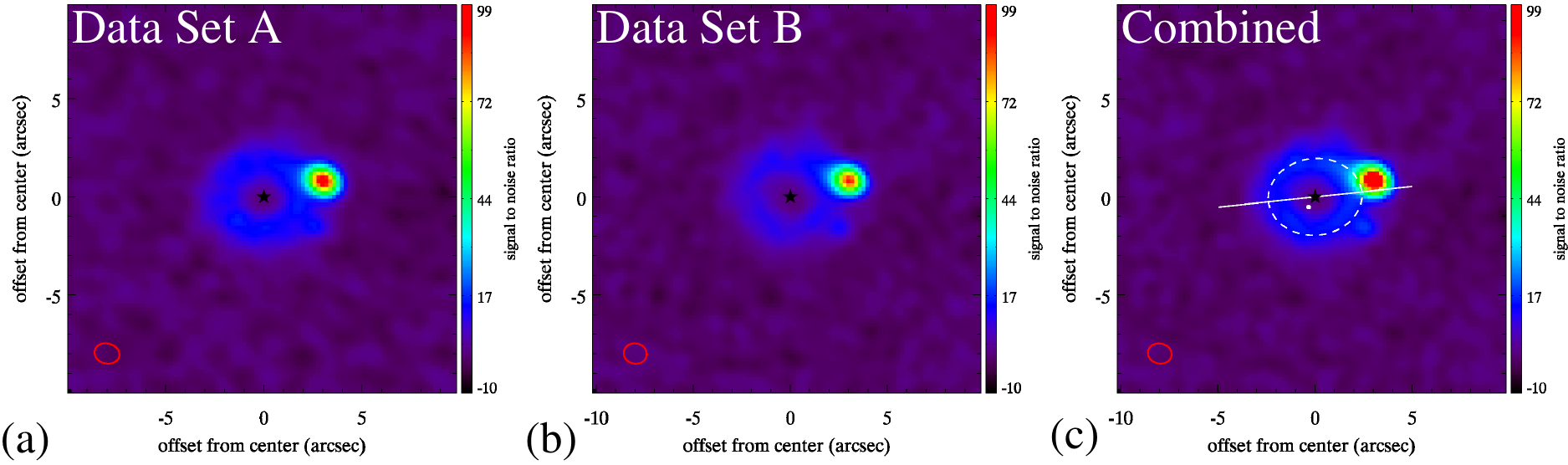}
  \caption{ALMA 1.3 mm continuum maps of the HD~95086 System. (a) and
(b) are the data obtained under program \#2013.1.00773S (PI: Su)
referred to as data set A, and program \#2013.1.00612S (PI: Booth)
referred to as data set B; (c) is the combined map using both data
sets. Details about the synthesized beams (shown as the red ellipse in
each of the panels) and rms of the maps are given in Sec 3.1.  In
panel (c), we also mark the positions of the star and its planet b (as
in 2016) as the black star symbol and white dot, respectively. The
ring's circumference (the dashed ellipse in panel (c)) is clearly
detected at S/N $\gtrsim$15 per beam. The bright source, detected at
S/N of 100, is almost aligned with the ring's major axis (white
line). Its nature is discussed in Section 4.3.}
\end{figure*}

\section{Observations}

We observed HD~95086 with ALMA in Band 6 (1.3 mm) under two projects:
\#2013.1.00773.S, PI: Su (referred to as data set A) and
\#2013.1.00612.S, PI: Booth (referred to as data set B). The observations
consist of 12 single pointing block executions centered at HD~95086
(phase center RA: 10:57:02.91 Dec: --68:40:02.27 (J2000)). The majority of
the observations were obtained in April/May 2015, while one was done
in January 2015. The proper motion of the star (pmra = --41.11$\pm$0.03
mas/yr and pmdec = 12.91$\pm$0.03 mas/yr) gives 11 mas offset for the three
month time span, i.e., there is no significant pointing difference in these
observations. Table \ref{tab1_obs} lists the details about these
observations including dates, block id, number of antennae used,
projected baselines, weather conditions, on-source integration time and
flux calibrators.

The correlator set-up was designed to optimize the continuum
sensitivity, but also covered the $^{12}$CO J=2-1 transition at
230.538 GHz with 3840 channels over a bandwidth of 1.875 GHz. The
set-up was slightly different between the two projects. The four
basebands were centered at 215, 217, 230 and 232.5 GHz for data set A,
but at 231.87, 232.55, 245, 247 GHz for data set B. The raw data were
processed by the ALMA Regional Centers using the CASA package (ver.\
4.2.2 for data set A and ver.\ 4.3.1 for data set B). Nearby quasars
and solar system objects (Callisto and Ganymede) were used for flux
calibration, resulting in an absolute flux uncertainty $\lesssim$10 \%
(the Technical Handbook for cycle 2). The total on-source integration
time is 4.58 hours for data set A, and 4.54 hours for data set B. No
CO detection was reported in the pipeline reduced product. Details
regarding the CO gas in the system will be reported in another
publication (Booth et al.\ in prep.).

\section{Results and Analysis}

\subsection{Continuum Emission}

We generated the calibrated measurement sets using the scripts
provided by the ALMA project for each of the data sets. We then split
the observations into different fields (pointing) and spectral windows
by binning the time sampling to 30 s and averaging the spectral
channels with a width of 128 channels. These averaged, binned $uv$
visibilities were then exported to FITS format for further analysis
using the MIRIAD software \citep{sault95}. Visibilities were then
inverted with natural weighting, deconvolved, and restored to generate
a final synthesized map using the standard procedures in MIRIAD.

For the data set A, the synthesized continuum image is shown in Figure
\ref{alma_images}a with a synthesized beam of
1\farcs28$\times$1\farcs03 and a position angle (P.A., measured from
North toward the East) of 75.1\arcdeg, and a rms of 8.7 $\mu$Jy
beam$^{-1}$. For the data set B, the image is shown in Figure
\ref{alma_images}b with a synthesized beam of
1\farcs20$\times$1\farcs04 and a P.A. of 79.4\arcdeg, and a rms of
11.0 $\mu$Jy beam$^{-1}$. In both images, a ring-like structure is
clearly seen with a very bright point-like source offset from the star
$\sim$3\arcsec\ away at a P.A. of 293\arcdeg\ (--67\arcdeg). Since the
quality (rms and beam) of both data sets was similar, we then combined
both data sets and generated a slightly deeper continuum map (shown in
Figure \ref{alma_images}c).  The combined continuum map has a
synthesized beam of 1\farcs22$\times$1\farcs03 and a P.A. of
77.4\arcdeg, and a rms of 7.5 $\mu$Jy beam$^{-1}$. The ring's
circumference is detected at signal-to-noise (S/N) $\gtrsim$15
$\sigma$ per beam, and is slightly inclined from face-on.  We estimate
the pointing accuracy of the data, $\sim$resolution/signal-to-noise,
to be 0\farcs13 since the main ring is detected at S/N$\gtrsim$10.

We adopt two approaches to explore the best-fit parameters for the 
HD~95086 system: (1) visibilities fitting and (2) image plane fitting.
In both approaches, we assume the millimeter emission can be described
by an optically and geometrically thin (no scale height) model plus a
point source offset from the center.  We explore two simple
axi-symmetric models to describe the disk: (1) a two-boundary disk
confined in a radial span of $R_{in}$ and $R_{out}$ with a surface
density power law of $\Sigma(r) \propto r^{p}$ where $r$ is the
stellocentric distance, and (2) a Gaussian ring defined by the peak 
($R_p$) and the width (FWHM) of the ring ($R_w$).  For the
millimeter emission of the disk (i.e., dominated by large grains), we
expect the dust temperatures follow $T_d(r) = 278.3 L_{\ast}^{0.25}
r^{-0.5}$ where $L_{\ast}$ is the stellar luminosity in units of the
solar luminosity (6 $L_{\sun}$ for HD~95086 using the new distance)
and $r$ is in au. The disk has a total flux, $F_{tot}$, at
1.3 mm, and its mid-plane is assumed to incline by an angle of $i$
from face-on (i.e., $i=0$\arcdeg) with the major axis along a position 
angle (P.A.). There are three parameters describing the bright source: the
total flux ($F_{pt}$) and offset from the star ($\Delta x$ and $\Delta
y$). We discuss the results in the following subsections for both
approaches, and synthesize the final best-fit model parameters in
Section 3.4.

\begin{figure*}[bht] 
  \figurenum{2}
  \epsscale{1.1}
  \label{mcmc_model}    
  \plotone{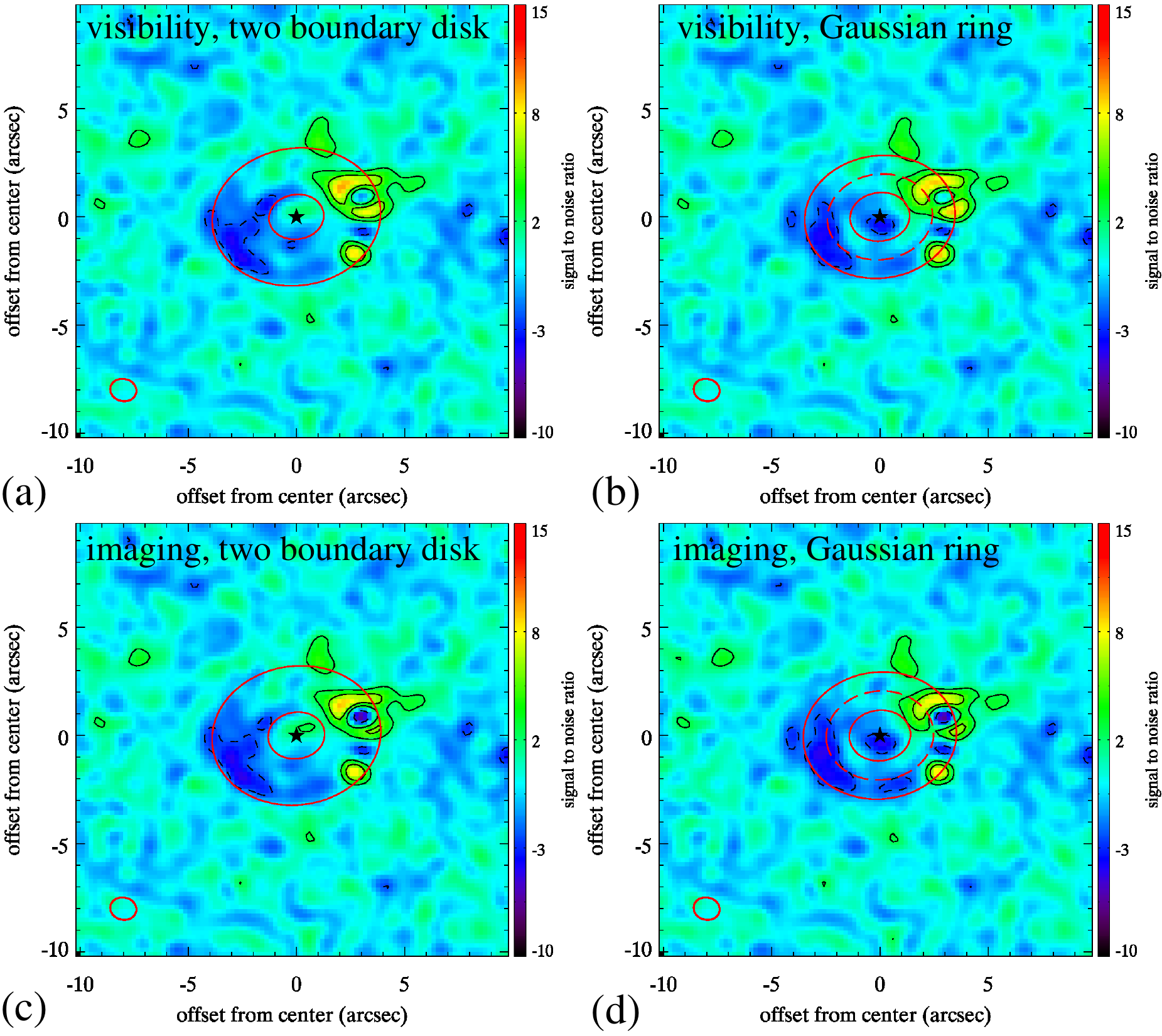}
  \caption{Residual maps of the HD~95086 system after subtracting four
best-fit models (see Table 2). Panels (a) and (b) are the results
using the visibility modeling approach while panels (c) and (d) are
from the image plane fitting. The left column is for the two boundary disk
model where the boundaries of the disk are marked with the two red
ellipses. The right column is based on the Gaussian ring model where
the peak of the ring is marked as the red dashed ellipse and the
boundaries of the ring ($R_p\pm 0.5R_w$) are also shown. The display
orientation (N up and E left), color scale, contours and the star
position (black star symbol) are all the same in each of the panels in
units of signal-to-noise with rms of 7.5 $\mu$Jy~beam$^{-1}$. The
contour levels are in [--3, 3, 6]$\times$rms. }
\end{figure*}

\begin{deluxetable*}{llccccccccccc}
\tablewidth{0pc}
\footnotesize 
\tablecaption{MCMC Derived Disk Parameters with One Point-like Bright Source \label{tab2_mcmc}}
\tablehead{
\colhead{Parameter$^1$ } &  \colhead{Description}  & \multicolumn{5}{c}{Two Boundary Disk} & & \multicolumn{5}{c}{Gaussian Ring} }
\startdata
                    &                         &\multicolumn{2}{c}{Visibilities modeling} & &\multicolumn{2}{c}{Dirty map modeling} & &\multicolumn{2}{c}{Visibilities modeling} & &\multicolumn{2}{c}{Dirty map modeling}  \\
                    &                         & Value & $\pm$1$\sigma$ & & Value & $\pm$1$\sigma$ & & Value & $\pm$1$\sigma$ & & Value & $\pm$1$\sigma$ \\
\cline{3-4} \cline{6-7} \cline{9-10} \cline{12-13} \\
$R_{in}$  [au]      & inner belt radius       &  107    &  +6 --5     &&  110    &  +3 --4     &&\nodata  & \nodata    &&\nodata  & \nodata      \\   
$R_{out}$ [au]      & outer belt radius       &  327    &  +6 --7     &&  328    &  +7 --6     &&\nodata  & \nodata    &&\nodata  & \nodata      \\   
$p$        & surface density index            & --0.48   & +0.34 --0.38&& --0.42  & +0.13 --0.12&&\nodata  & \nodata    &&\nodata  & \nodata      \\
$R_{p}$   [au]      & peak radius             &\nodata  & \nodata     &&\nodata  & \nodata     &&  204    & +7 --7     &&  208    & +4 --3       \\   
$R_{w}$   [au]      & width (FWHM)            &\nodata  & \nodata     &&\nodata  & \nodata     &&  176    & +6 --6      &&  179    & +6 --6       \\   
$F_{tot}$ [mJy]     & total belt flux density &  2.87   & +0.10 --0.11&&  2.89   & +0.08 --0.08&&  2.91   & +0.10 --0.18&&  3.07   & +0.09 --0.09 \\   
$F_{pt}$ [mJy]     & flux of Pt.$^2$          &  0.88   & +0.05 --0.05 &&  0.92   & +0.01 --0.01 && 0.87   & +0.06 --0.06 &&  0.92   & +0.01 --0.01 \\   
$\Delta$x [\arcsec] & RA offset of Pt.     &  --3.06   & +0.04 --0.04 & &  --3.07   & +0.01 --0.01 &&  --3.05   & +0.05 --0.05 &&  --3.07   & +0.01 --0.01 \\   
$\Delta$y [\arcsec] & Dec offset of Pt.    &  0.85   & +0.05 --0.05 &&  0.85   & +0.01 --0.01 &&  0.85   & +0.05 --0.05 &&  0.85   & +0.01 --0.01 \\   
$i$       [\arcdeg] & inclination             &  36   & +3 --2   &&  35   & +2 --2   &&  36   & +2 --2   &&  34   & +2 --2   \\   
P.A.      [\arcdeg] & position angle          &  98   & +3 --3   &&  98   & +3 --3   &&  98   & +3 --4   &&  96   & +4 --3   \\   
\enddata 
\tablenotetext{1}{We adopt a distance of 83.8 pc to translate the angular scale to physical scale.}
\tablenotetext{2}{Pt. is the bright point-like source near the edge of the disk.}
\end{deluxetable*}

\subsection{Visibilities Modeling Approach}

We model the visibilities for both data sets simultaneously.  To
minimize the free parameters, we assume no offset between the center
of the disk and the star. Therefore, there are a total of eight/nine free
parameters to describe the system in both axi-symmetric disk models:
two/three parameters for the disk density distribution ($R_{in}$, 
$R_{out}$ and $p$ for the two-boundary disk, or $R_{p}$ and $R_{w}$ for the
Gaussian ring), two parameters for the disk viewing geometry ($i$ and
P.A.), the total flux of the disk ($F_{tot}$), and three parameters
for the point source ($\Delta x$, $\Delta y$ and the total flux
$F_{pt}$). We determine the best-fit values for these free
parameters independently by adopting the MCMC approach outlined in
\citet{macgregor13}.  For all parameters, we assume uniform priors and
require that the model be physically plausible (flux greater than zero and 
the outer radius larger than the inner one).

The best-fit parameters and their $\pm$1$\sigma$ uncertainties are
given in Table \ref{tab2_mcmc}. For each set of the best-fit
parameters, we generated a high resolution model image and transformed
it to the visibility domain according to the observation. We then
constructed the residual map by subtracting the model from the data in
the visibility domain and imaging the residual using the same
procedures in MIRIAD. The residual maps are shown in Figure
\ref{mcmc_model}.  Overall, the residuals are within $\pm$3$\sigma$
for the main disk. The subtraction of the bright source is not
perfect, and creates an over-subtraction at the center of the bright
source, and positive residuals in the area around it, suggesting the
source might be extended . In all residual maps, there appears to be
another faint (S/N$\sim$9) source $\sim$2\farcs5 south of the bright one. The two
axi-symmetric models yield very similar parameters in terms of the
disk flux, viewing geometry and point source parameters. However, the
residual in the main disk is slightly smaller in the Gaussian ring
model. Although the residuals in the main disk tend to be more
negative in the east side of the disk, unfortunately the bright source
is along the west side of the major-axis, making it difficult to
assess any asymmetric structure present in the disk (more detail is
given in Section 4.2).

\subsection{Imaging Plane Modeling Approach} 

Given the good S/N detection of the main disk, we also try to derive
the best-fit parameters for the two models by fitting in the image
plane. Details about this approach can be found in \citet{booth16}. We
use the combined synthesized map for the MCMC search.  For
experiments, two more free parameters are included in this part of the
fitting.  For both models, the center of the ring is not fixed at the
star position. Although a small offset\footnote{0\farcs16 and
0\farcs06 for the two boundary model, and 0\farcs21 and 0\farcs08 for
the Gaussian ring} is preferred for both models, these values are
within one pixel of the reconstructed maps (0\farcs2 per pixel) and within 2
$\sigma$ of the pointing accuracy ($\sigma\sim$ 0\farcs13), therefore
not significant. The final best-fit parameters and the residual maps
are also given in Table \ref{tab2_mcmc} and Figure
\ref{mcmc_model}. Overall, the best-fit parameters agree with the ones
derived from the visibilities method within the uncertainties, except
for the total fluxes of the ring and the point source where the
derived flux using the imaging approach is consistently larger. We
also note that the estimated uncertainties are also smaller using the
image plane approach. This is because the MCMC uncertainty depends
strongly on the weightings of the data. The noise within the beam is
highly correlated, and the image deconvolution (the ``CLEAN''
procedure) treats noise non-linearly, both resulting in smaller
uncertainties for the MCMC image fitting that might not be statistically
robust. A factor equivalent to the square root of the beam size in pixels
is included to mitigate the correlated noise, but this is only an 
approximation as it assumes a Gaussian beam whereas the dirty beam has
some low level, non-Gaussian structure that this factor cannot account for. 
Fitting in the image domain is computationally faster, and can
achieve the same result in terms of geometric parameters for high S/N
data; however, we caution against relying on the robustness of the
uncertainties using imaging plane fitting.

\begin{deluxetable}{clrr}
\tablewidth{0pc} 
\tablecaption{Derived Parameters of the two Point Sources \label{tab3_twops}}
\tablehead{\colhead{Parameter } &  \colhead{Description}   &  \colhead{Point 1}  &    \colhead{Point 2} }
\startdata
$F_{pt}$ [mJy]     & flux density  &  0.81  $\pm$0.03 &  0.10$\pm$0.02  \\   
$\Delta$x [\arcsec] & RA offset     &  --3.08  $\pm$0.04 &  --2.80$\pm$0.03  \\   
$\Delta$y [\arcsec] & Dec offset    &  0.83  $\pm$0.05 & --1.61$\pm$0.04 
\enddata
\end{deluxetable}

\begin{figure*}[htb] 
  \figurenum{3}
  \epsscale{1.1}
  \label{pts_ring}    
  \plotone{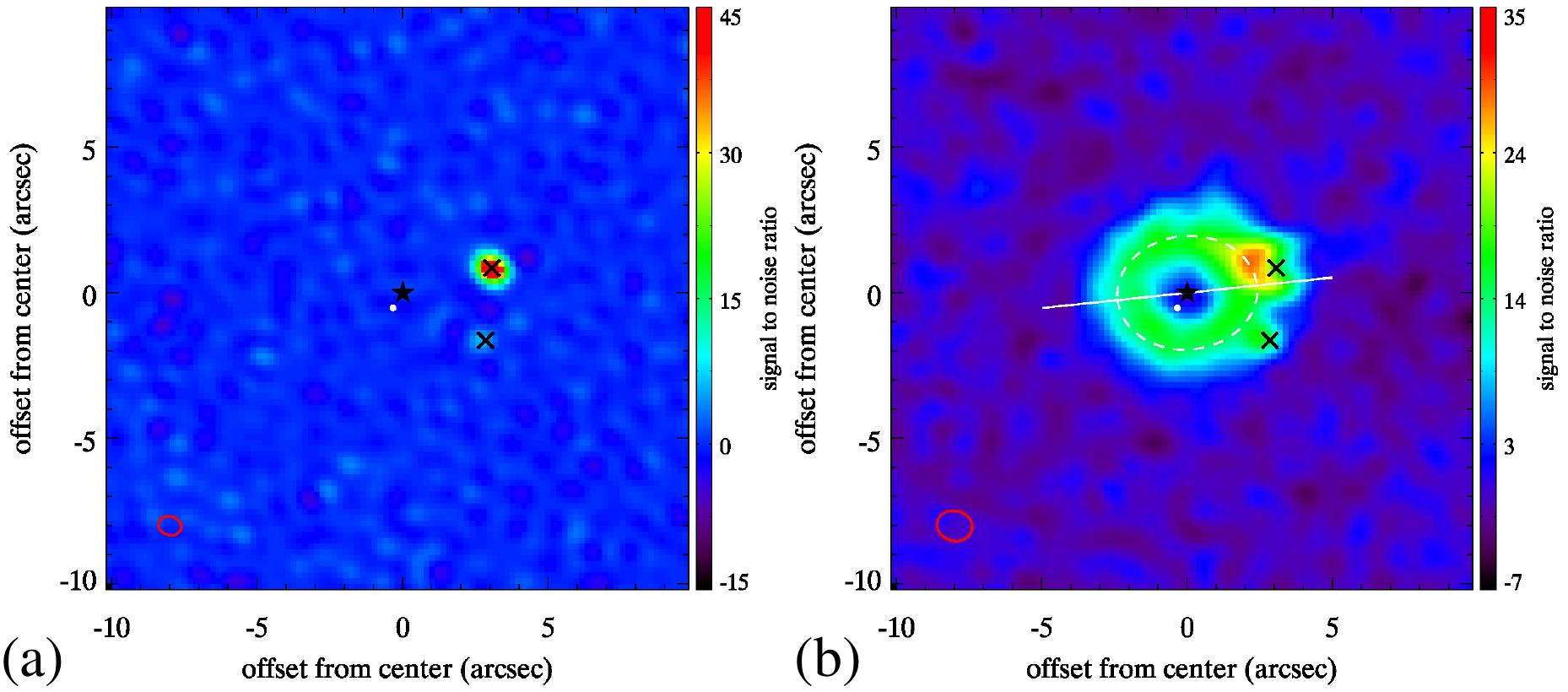}
  \caption{1.3 mm continuum map of the point sources in (a) and the
disk in (b).  The left panel is generated using the long baseline data
($>$80 $k\lambda$) to better illustrate the bright point-like sources
(marked as black crosses). The rms in the long baseline map is 11.4
$\mu$Jy~beam$^{-1}$ with its synthesized beam shown as the red
ellipse. The right panel is the ``disk-only'' map generated after the
subtraction of the two point sources.  The positions of the star
(black star symbol), planet b (white dot) and disk major axis (white
line) are also marked.}
\end{figure*}

\begin{deluxetable*}{llccccccccccc}
\tablewidth{0pc}
\footnotesize 
\tablecaption{MCMC Derived Disk Parameters by Minimizing Other Contamination$^1$ \label{tab4_disk}}
\tablehead{
\colhead{Parameter$^2$ } &  \colhead{Description}  & \multicolumn{5}{c}{Two Boundary Disk} & & \multicolumn{5}{c}{Gaussian Ring} }
\startdata
                    &                         &\multicolumn{2}{c}{Visibilities}& &\multicolumn{2}{c}{Dirty map} & &\multicolumn{2}{c}{Visibilities } & &\multicolumn{2}{c}{Dirty map }  \\
                    &                         & Value & $\pm$1$\sigma$ & & Value & $\pm$1$\sigma$ & & Value & $\pm$1$\sigma$ & & Value & $\pm$1$\sigma$ \\
\cline{3-4} \cline{6-7} \cline{9-10} \cline{12-13} \\
$R_{in}$  [au]      & inner belt radius       &  105    &  +5 --5    &  &  106    &  +5 --4     & &\nodata  & \nodata   &   &\nodata  & \nodata      \\   
$R_{out}$ [au]      & outer belt radius       &  331    &  +4 --6    &  &  312    &  +7 --8     & &\nodata  & \nodata   &   &\nodata  & \nodata      \\   
$p$        & surface density index            & -0.5   & +0.3 --0.3 & & --0.5   & +0.2 --0.2    & &\nodata  & \nodata     &   &\nodata  & \nodata      \\
$R_{p}$   [au]      & peak radius             &\nodata  & \nodata     & &\nodata  & \nodata     & &  199    & +6 --6    &   &  200    & +4 --4       \\   
$R_{w}$   [au]      & width (FWHM)            &\nodata  & \nodata     &  &\nodata  & \nodata    &  &  169    & +5 --6   &    &  167    & +7 --7       \\   
$F_{tot}$ [mJy]     & total belt flux density &  2.77   & +0.19 --0.07 & &  2.57   & +0.09 --0.08 & &  2.72   & +0.12 --0.08 & &  2.74   & +0.09 --0.10 \\   
$\Delta$x$_c$ [\arcsec] & RA offset of ring     &  0.15   & +0.04 --0.04 & &  0.09   & +0.03 --0.03 & &  0.15   & +0.05 --0.04 & &  0.12   & +0.04 --0.04 \\   
$\Delta$y$_c$ [\arcsec] & Dec offset of ring     &  0.05   & +0.04 --0.04 & &  0.05   & +0.03 --0.03& &  0.06   & +0.04 --0.04 & &  0.08   & +0.03 --0.03 \\   
$i$       [\arcdeg] & inclination             &  31   & +3 --3   &  & 28   & +3 --3 &  &  31   & +2 --3  & &  30   & +3 --3   \\   
P.A.      [\arcdeg] & position angle          &  98   & +4 --3   &  & 97   & +6 --6 &  &  98   & +4 --4  & &  95   & +5 --5   
\enddata 
\tablenotetext{1}{``Contamination'' mainly means the bright source near the edge of the disk. In the visibility approach, a Gaussian-like profile is used to fit the bright source. In the image plane fitting, the area affected by the bright source is masked out. Details see Section 3.4.} 
\tablenotetext{2}{We adopt a distance of 83.8 pc to translate the angular scale to physical scale.}
\end{deluxetable*}  

\subsection{Best-Fit Disk Parameters by Minimizing Other Contamination}

Since ALMA is sampling the sky with many different baselines (i.e.,
spatial scales) through interferometry, we can better assess the
properties of the bright source by generating a map with only the long
baseline data ($>$80 $k\lambda$ = 3\farcs1) where any extended
structure with sizes $>$3\farcs1 (i.e., the disk emission) is filtered
out. The long baseline map is shown in Figure \ref{pts_ring}a. Within
3\farcs5 radius from the star, there are two sources (detected above 8
$\sigma$) that appear in the long baseline map. The measured FWHM of
the faint source is 0\farcs79$\times$0\farcs64, the same as the
synthesized beam in the long baseline map, while the FWHM of the
bright source is $\sim$8\% broader, 0\farcs85$\times$0\farcs70. To
evaluate whether the threshold defining the long baseline data affects the
FWHM of the bright source, we also generated the long baseline maps
with different thresholds between 60 to 80 $k\lambda$. The FWHM of the
bright source is consistently broader than the synthesized beam by
8\%. Furthermore, we generated individual, long baseline maps per data
set to see if there exists a flux difference between the two data sets
for the bright source. Taken at face value, the bright source is
about 5\% brighter in data set B. Given the typical absolute flux
uncertainty (10\%) in the ALMA data, the flux difference is not
significant. The properties (offset and flux density) of the two point
sources are determined using 'uvmodelfit' in CASA, and given in Table
\ref{tab3_twops}. Using the long baseline data, the flux of the bright
source is 9\% lower than the derived flux by fitting the disk and
point source simultaneously.

\begin{figure*}[htb] 
  \figurenum{4}
  \epsscale{1.1}
  \label{residual_disk}    
  \plotone{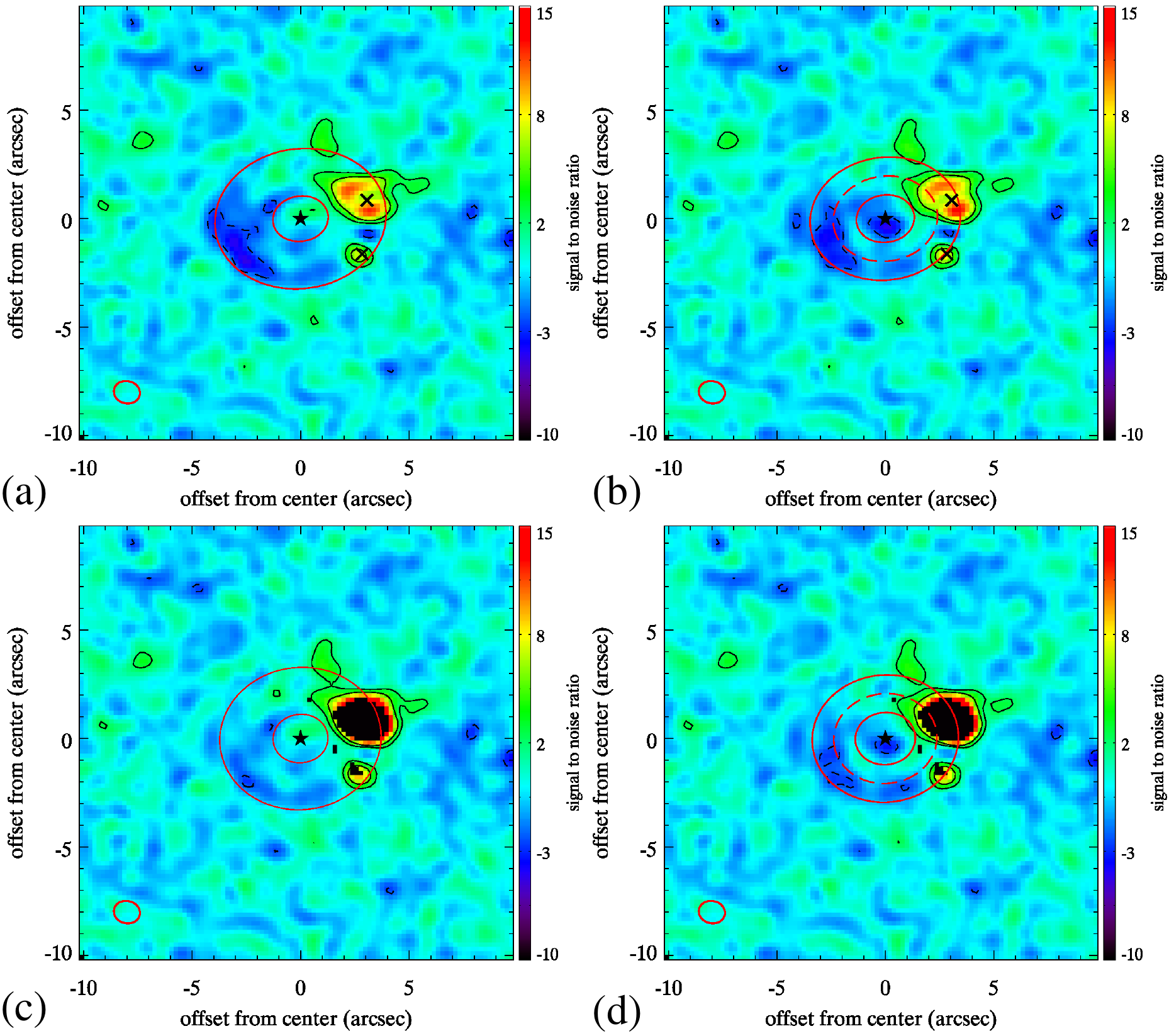}
  \caption{1.3 mm residual maps using the best-fit parameters derived
in Section 3.4 by minimizing the contamination from the bright source.
(a) and (b) are the residual maps for the two boundary disk and the
Gaussian ring obtained by fixing the two point source parameters in the
visibility fitting. (c) and (d) are the similar residual maps using
the image plane fitting by masking pixels (black area) affected by
the bright, point-like sources (details see Section 3.4). Other lines, 
contours and symbols are the same as in Figure \ref{mcmc_model}.}
\end{figure*}

We then generated a ``disk-only'' map by subtracting the best-fit point
sources in the visibility domain. The ``disk-only'' map is shown in
Figure \ref{pts_ring}b. The subtraction of the bright source is far
from perfect. There is still significant flux ($\gtrsim$25 $\sigma$)
near the east side of the bright source, which could be part of the
disk structure or the bright source has a non-symmetric, extended
shape. If the bright source is a dusty galaxy in nature (see Section
4.3), it is very likely to have an irregular shape, making it
challenging to separate it from the disk without high angular
resolution observations. The subtraction of the faint source is
better. The emission near the position of the faint source is more
smooth, but it does appear that the disk flux extends toward the faint
source, explaining the residual in Figure \ref{mcmc_model}. Similarly, 
the observed resolution prevents further assessment.

To evaluate the impact of the two sources on the
derived disk parameters, we searched for the best-fit disk parameters
using the visibility approach by fixing the properties of the two
point sources.  The best-fit disk parameters are basically the same
as the ones without fixing the two point sources (Section 3.2). The
residual maps are shown in Figures \ref{residual_disk}a and
\ref{residual_disk}b. Compared to Figures \ref{mcmc_model}a and
\ref{mcmc_model}b, the fits with the fixed point source properties
have no over subtraction at the position of the bright source, but the
residual around the bright source is higher. 

We performed similar searches using the image plane approach by
masking out the pixels that have fluxes $>$0.1 mJy~beam$^{-1}$ (the
region around the bright source and the center of the faint one). To explore
whether allowing a slight offset between the star and the disk center
can improve the results, we also allow offsets in the MCMC parameters.
The best-fit disk parameters are given in Table
\ref{tab4_disk}. Interestingly, the disk size parameters are slightly
smaller than the ones derived in Section 3.3, but still within
uncertainties. The residual maps are shown in Figures
\ref{residual_disk}c and \ref{residual_disk}d.  The derived disk
fluxes is also smaller, reducing the over-subtraction in the east side
of the disk. The presence of the bright source is undoubtedly
affecting the derived disk parameters. Unfortunately, there is no easy
way to mask out the contribution of the bright source in the
visibility domain since the bright source contributes to all
baselines. We tried to minimize the impact of the bright source by
fitting it as a Gaussian profile in visibilities, allowing for some
sort of extension. The results are also given in Table
\ref{tab4_disk}. Similar to the image plane approach, the disk size
parameters are also slightly smaller than the ones derived in Section
3.2. The Gaussian parameters for the bright source are basically the
same as the synthesized beam, but with much less flux compared to
the fits derived from the long baseline data (Table
\ref{tab3_twops}). The residual maps (not shown) are neither better
than the ones in Figures \ref{residual_disk}a and \ref{residual_disk}b
nor than those in Figures \ref{mcmc_model}a and \ref{mcmc_model}b as a result.

We synthesized the various fitting results as follows. The disk has a
total flux density of 2.79$\pm$0.1 mJy at 1.3 mm, and is inclined by
30\arcdeg$\pm$3\arcdeg\ from face-on with a P.A. of
97\arcdeg$\pm$3\arcdeg. It is interesting to note that the best-fit
models (Table \ref{tab4_disk}) all prefer to have the ring center east
of the star by $\sim$0\farcs1. Since the star is not detected in the
ALMA data, the ``translated'' pointing accuracy cannot confirm such an
offset. The disk is very broad ($\Delta R/R\sim$0.84) in the
millimeter continuum, and its width is resolved by $\sim$1.7 beam
widths. As a result, we cannot determine the exact disk density
distribution nor the offset between the star and disk center. For the
two-boundary model, the disk can be described as having sharp
boundaries at $R_{in}=$106$\pm$5 au and $R_{out}=$320$\pm$10 au with a
surface density power index of --0.5$\pm$0.3. For the Gaussian ring, the disk
is peaked at $R_p=$200$\pm$6 au with a width of $R_w=$168$\pm$7 au.
The reduced chi-squared ($\chi^2$) is 1.20 for the two-boundary model,
but 1.15 for the Gaussian ring model with the total number of
visibilities (574358) in the fitting; similar $\chi^2$ numbers are
also found using the image plane approach. Based on the $\chi^2$, the
Gaussian ring model gives a slightly better fit; however, the number
of free parameters is different (7 vs.\ 8), and some of the parameters
are correlated.

\section{Discussion}

\subsection{Revised 3-Component SED Model} 

\begin{figure}[htb] 
  \figurenum{5} 
  \label{sed} 
  \plotone{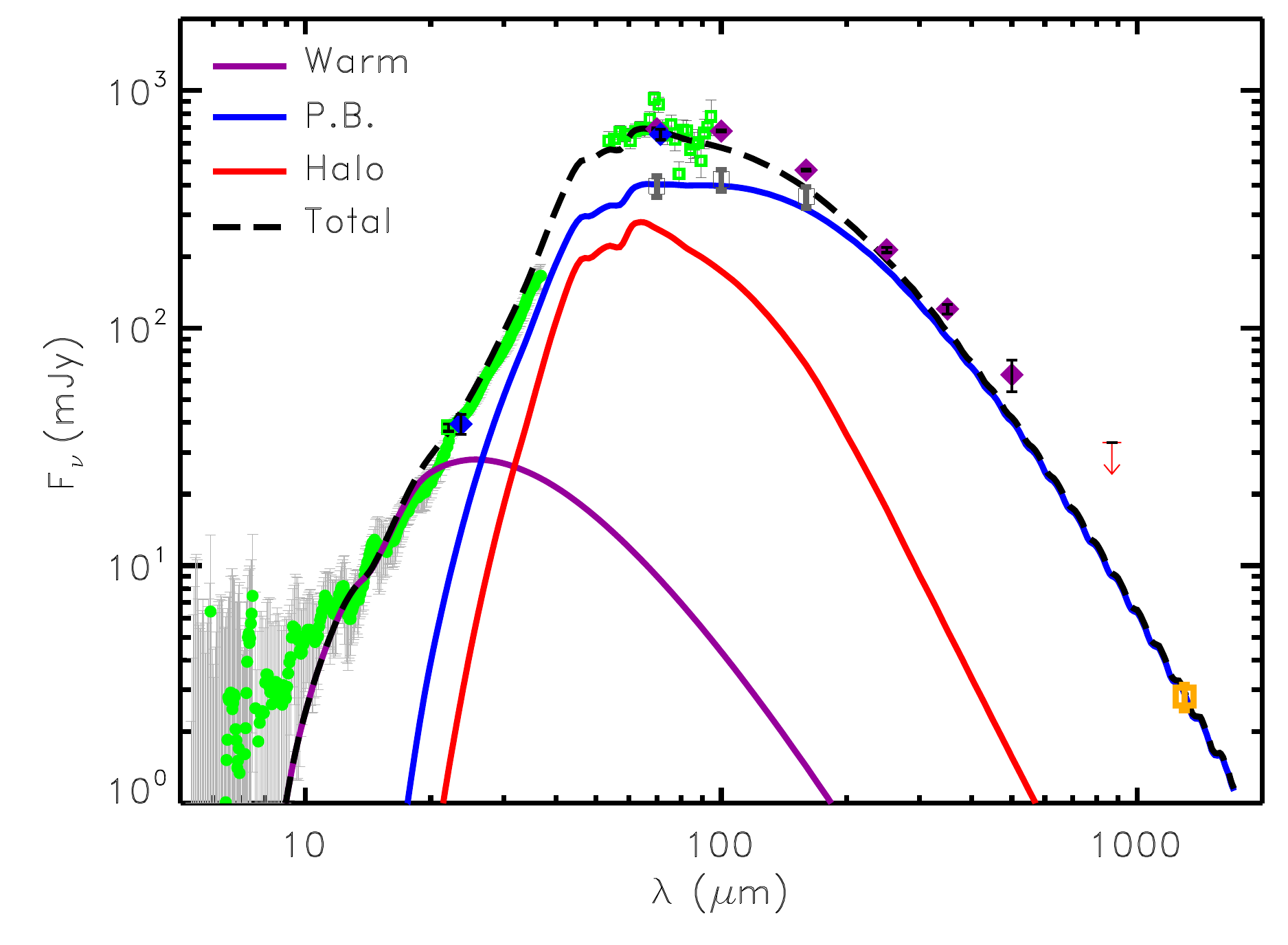}
  \caption{Revised 3-component SED model of the HD~95086
system. Various symbols with error bars are measurements from
\citet{su15}. The estimated 1.3 mm disk flux from this study is shown
as the orange square. Various lines are the SED models for the warm,
planetesimal (P.B.) and disk halo components. The photometry points
from {\it Herschel} (purple diamonds) and APEX are likely contaminated
by the point source detected in our ALMA map, therefore, the total
model disk SED (dashed line) is lower than those values. The grey
squares are the maximum allowable fluxes derived from the PACS data
(details see \citealt{su15}).}
\end{figure}

\citet{su15} examined the resolved disk images of HD~95086 from {\it
Herschel} and argued, by a detailed SED analysis, that the system is 
likely to possess three debris
structures. The three debris
components are very similar to the ones of the HR 8799 system -- (1)
an inner warm emission representing the dust in an asteroid-belt
analog, (2) an outer cold emission representing the dust in a
Kuiper-belt analog, and (3) an extended disk halo surrounding the
aforementioned two components and composed of small grains. Since only
the extended disk halo is resolved in the far infrared, the exact
boundaries of the different components are very uncertain and not well
constrained by the SED model. The presence of an asteroid-belt analog
is only constrained by the excess emission detected in the {\it
Spitzer} IRS spectrum and unresolved MIPS 24 \um photometry;
therefore, its location is set by the observed dust temperature
($\sim$175 K, i.e., 7--10 au). Given the warm temperature and large
distance to the system, this asteroid-belt component is not expected
to be detected nor resolved by the ALMA observation.

Since we now have the measured size for the cold disk ($R_{in}$=106 au
and $R_{out}$=320 au compared to the old SED 63--189 au value), the
3-component SED model needs revision. Furthermore, a re-reduction of
the archive APEX/LABOCA 870 \um data on HD~95086 using the techniques
described in \citet{phillips11} finds a total flux of 19.4$\pm$11.0
mJy, much lower than the one published by \citet{nilsson10}. The peak
emission in the 870 \um map is offset by 13\arcsec\ (i.e., 2/3 of the
beam diameter) from the expected star position; therefore, the quoted
flux is estimated as an unresolved source at the position of the
star. Due to the large offset, we consider it as a non-detection and
use 33 mJy as the 3 $\sigma$ upper limit. We adopted this value for
further SED analysis. Using similar approaches and grain parameters as
in \citet{su15} (minimum and maximum grain sizes of $\sim$1.8 \um and
1000 \mm, and a particle size distribution in a power law index of
$-$3.5), a geometrically thin, constant surface density disk with a
radial span of 106-320 au provides a good fit to the ALMA 1.3 mm flux
and maximum allowable fluxes in the far infrared (gray squares in
Figure \ref{sed}). This cold disk SED model also agrees with the
observed 7 mm flux (not shown in Figure \ref{sed}) obtained by
\citet{ricci15b} within the uncertainty.  Compared to the cold
component model presented in \citet{su15}, this revised planetesimal
disk model contributes much more flux shortward of 60 \mm, and is the
dominant component (compared to the disk halo) at 20--30 \mm. With a
much larger cold planetesimal disk component, the inner radius of the
disk halo is more distant from the star (from old $\sim$190 au to
$\sim$300 au), and contributes much less flux at mid- and far-infrared
wavelengths using the same grain parameters as in
\citet{su15}. Although we now have resolved the cold disk at
millimeter wavelengths, revealing the placement of the large grain
population, there are still a wide range of parameters that are not
constrained in the SED models, especially for the grain parameters in
the disk halo. The SED models shown in Figure \ref{sed} are not
unique. Future resolved images of the various components at crucial
mid-infrared wavelengths will shed light on this.

\subsection{Ring Width and Azimuthal Asymmetry}

\begin{figure}[htb] 
  \figurenum{6} 
  \label{diskprofiles} 
  \plotone{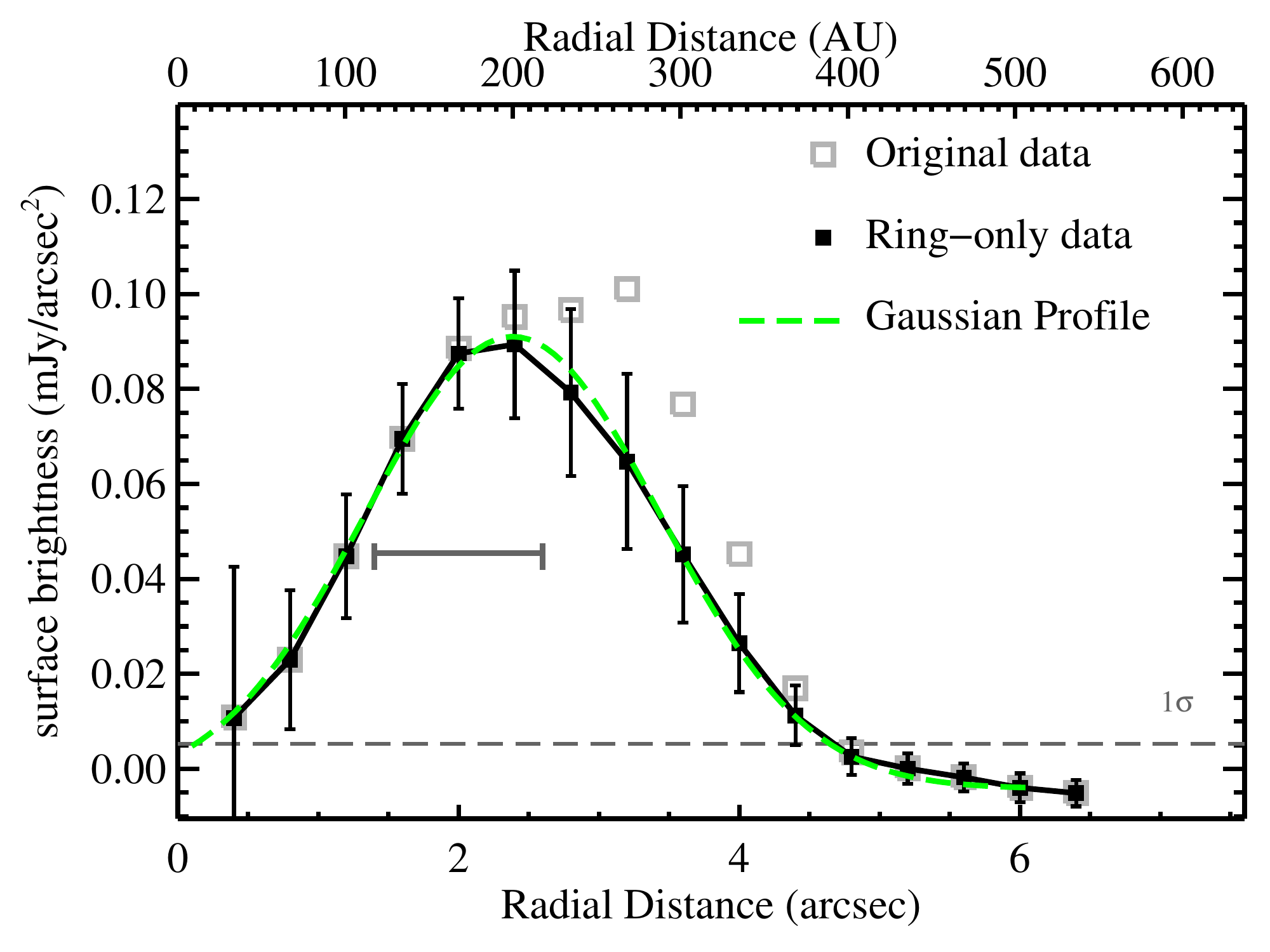}
  \plotone{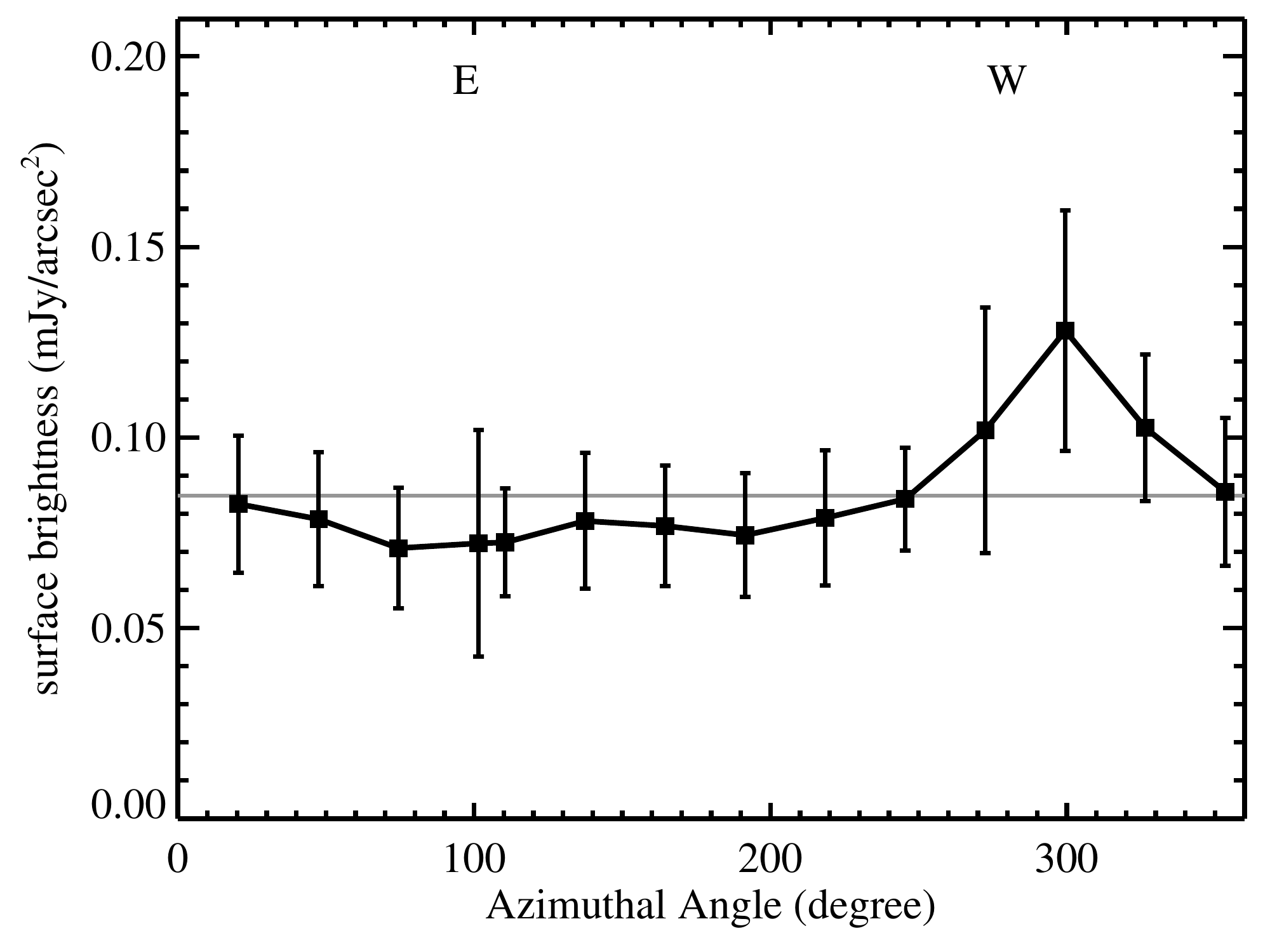}
  \caption{The top panel shows the azimuthally averaged radial surface
brightness profile of the disk. Filled, connected symbols are the
profiles using the disk image after the subtraction of the two point
sources, while the grey symbols are the one without. A best-fit,
symmetric Gaussian profile is shown as the solid green line for
comparison. The horizontal dashed line shows the rms of the images,
and the grey bar marks the radial span that was used to compute the
azimuthal angle profile. The bottom panel is the surface brightness
profile of the disk along the disk circumference after the subtraction
of the two point sources. The horizontal grey line represents the
average disk surface brightness. }
\end{figure}

We computed the azimuthally averaged radial profile for the disk using
the synthesized image data without the two point sources (the right
panel of Figure 3).  Assuming the disk is inclined by 30\arcdeg\ with
the major axis along P.A. of 97\arcdeg, we first created a series of
elliptical rings with a width of 2 pixels (0\farcs4) centered at the
star, and computed the average value of all pixels that fall in each
ring. Since the pixels are highly correlated within the area of each
synthesized beam, the noise in each ring can be approximated with the
standard deviation in the ring divided by the number of beams in that
ring. The background noise per ring is computed in a similar
fashion. The total error in the average flux measurement per ring,
therefore, is the nominal error and the background noise added in
quadrature. The resultant disk surface brightness profile is shown in
the top panel of Figure \ref{diskprofiles}. For reference, the profile
using the original data is also shown, and the contamination from the
bright source obviously results in extra flux in the radii at
2\arcsec--4\arcsec\ from the star. The point-source subtracted ring
profile is centered at $r\sim$2\farcs3 from the star with symmetric
profiles inside and outside the peak within uncertainties. The
millimeter emission of the disk is surprisingly broad ($\Delta
R/R\sim$0.84).

To assess the degree of the asymmetry in the millimeter disk emission,
we also computed the radially averaged surface brightness profile
along the disk circumference. To minimize the contamination of the
bright source, we picked a radial span of 1\farcs4--2\farcs6 from the
star, and computed the average disk brightness within an incremental
angle of 27\arcdeg\ azimuthally. Similarly, the uncertainty includes
the standard deviation and background noise in each of the wedges. The
azimuthal profile is shown in the bottom panel of Figure
\ref{diskprofiles}. The azimuthal profile agrees within 1 $\sigma$ to
the average disk surface brightness (horizontal grey line in Figure
\ref{diskprofiles}), except at P.A. of $\sim$300\arcdeg, the direction
toward the bright source (although the difference is still within 2
$\sigma$). Given the contamination by the bright source, plus the
modest resolution of the ring ($\sim$6 beam widths in circumference 
and $\lesssim$2 beam widths in width), the apparent asymmetry is not
significant.

\subsection{Possible Nature of the Bright Source} 

As demonstrated in Section 3.4, the bright source near the edge of the
disk, at 3\farcs2 (a projected distance of 268 au) from the star, is
slightly more extended than the synthesized beam, and roughly along the
major axis of the disk (3\arcdeg\ off). Although the disk emission can
be traced up to $\sim$320 au from the star, the peak millimeter
emission is within 200 au, i.e., this bright structure is quite far
away from the main location of the colliding planetesimals. We
explore various possibilities for the nature of the bright source
either physically associated with the HD~95086 system or due to chance
alignment of a background source.

\subsubsection{Debris Phase of a Circumplanetary Disk?}

It is challenging to form planets at large orbital distances through
the usual route of km-sized planetesimal merger. However, there are
multiple ways to bypass this hurdle like pebble accretion in
conjunction with planet scattering (e.g., \citealt{lambrechts12,
kenyon15,bromley16}). Therefore, it may be possible to have a newly
formed planet at $\sim$270 au from this 17 Myr old star.  Given the
fact that there is no sign of this bright source in the deep $K/L'$
data (J. Rameau priv.\ comm.) and in the mid-infrared photometry of
the system \citep{moor13,su15}, it is unlikely that the bright source
is the direct detection of a newly formed planet\footnote{The typical
temperature for a protoplanet is expected to be a few 100 to a few
1000 K \citep{zhu15,eisner15}.}. Assuming the bright source is at the
distance of 83.8 pc, the bulk of the 1.3 mm flux suggests a bolometric
luminosity of 2$\times10^{-3}$ to 7$\times10^{-5} L_{\sun}$ assuming
it has a temperature of 30--100 K, which translates to a radius of
0.7--1.4 au for the optically thick emitting area. It is interesting
to note that the Hill radius for a 10 $M_{\oplus}$ planet at 270 au
around 1.6 $M_{\sun}$ star is $\sim$5 au. Therefore, the mm flux of
the bright source could come from the dust emission of a
circumplanetary disk (CPD) whose typical size is expected to be one
third of the Hill radius \citep{martin11,zhu15}.

The CPD is expected to be gas-rich around a planet in formation, like
a scaled-down version of a protoplanetary disk around a young star.
With a total flux of 0.81 mJy at 1.3 mm, the estimated dust mass is
$\sim$0.2--0.4 $M_{\oplus}$ assuming a typical dust opacity,
$\kappa_{1.3mm}$= 2.3 cm$^2$g$^{-1}$ \citep{beckwith90} and dust
temperatures of 30--60 K.  We do not detect any CO gas emission from
the bright source.  The noise level in the integrated CO (2-1) line
flux is 1.5$\times10^{-23}$ W~m$^{-2}$ assuming a velocity dispersion
of 4.6 km~s$^{-1}$ (twice the Keplerian velocity at 270 au) (Booth et
al.\ in prep.). The CO gas mass for the hypothesized CPD is less than
2.3$\times10^{-6} M_{\oplus}$ (1 $\sigma$, details see Booth et al.\
in prep.), suggesting an extremely low gas-to-dust mass ratio. If the
dust emission did come from the CPD of a newly formed planet, the CPD
might be in the ``debris'' phase as moons/satellites are being
formed. However, the mass fraction between the hypothesized CPD and
newly formed planet is uncomfortably high ($\sim$10$^{-2}$ for a
Neptune-size planet) in comparison to the typical mass fraction of
$10^{-4}$ between the satellites and the giant planets in the solar
system, making this ``debris CPD'' hypothesis unlikely.

\subsubsection{A Dust Clump due to A Giant Impact?}

Alternatively, a recent giant impact in the disk can create a bright,
concentrated region in the disk \citep{telesco05}. Depending on the
impact velocity, the disk morphology could remain in the
clump-dominated phase that lasts for a few orbital periods after a
giant impact \citep{jackson14}. Given the system's young age,
observing such a large impact at 270 au is not impossible. Next, we
estimate whether the brightness/mass in the clump is consistent with
such a scenario. Using the parameters derived in Section 3.4, the
bright source contributes $\sim$25\% of the total disk flux at 1.3 mm,
which is significantly larger than the clump in the $\beta$ Pic
disk ($<$ 4\% of the disk flux).

At 270 au, the orbital velocities are so low that most collisions
between large objects would be mergers\footnote{The typical impact
velocity is roughly the orbital velocity, which is $\sim$2.3
$km s^{-1}$ at 270 au around HD~95086. Given the escape velocity of 25
$km s^{-1}$ for a Neptune-like planet, a typical impact between two
Neptune-like objects at 270 au belongs to the merging collision
outcome based on the work by \citet{leinhardt12}.}  and not produce
much debris (with masses $\sim $ a few \% of the impactors,
\citealt{jackson14}).  To produce the amount of dust observed in the
bright source, objects with masses $\gtrsim$10--20 $M_{\oplus}$
(Neptune mass) are required. Impacts involving such large objects are
likely to be very explosive, i.e., the resultant clump is expected to
smear and spread very rapidly after the impact (a few orbital
periods). The fact that the bright source is relatively compact
($\sim$8\% broader than the synthesized beam in the long baseline
data) disfavors the origin of a giant impact.

\subsubsection{A Dust Clump due to Planetesimals trapped by an Unseen Planet?}

In addition, a concentrated dust clump can also be created by the
intense collision among the planetesimals trapped in the resonance
with an unseen planet \citep{wyatt03}, as one of the proposed origins
for the dust clump in the $\beta$ Pic disk.  The dust clump in the
$\beta$ Pic disk is also found to be very bright in CO gas emission,
probably released by the icy planetesimals \citep{matra17}. Therefore,
we might also expect to detect a significant amount of CO gas
associated with the bright source in HD~95086 if all the planetary systems
have a similar composition. The upper limit on the integrated CO (2-1)
line flux is $\sim$100 times fainter than the integrated line flux of
CO (2-1) line in the $\beta$ Pic disk after scaling by the distance
difference, while the dust flux in the clump is much brighter in the
HD~95086 disk. Given these comparisons, it seems unlikely the bright source
has a similar nature as the clump in the $\beta$ Pic disk.

\subsubsection{Alignment of a Background Galaxy} 

An alternative explanation is that the source is a background
galaxy. In fact, \citet{su15} suspected that the integrated
submillimeter flux of the system is likely contaminated by background
galaxies due to the excess emission detected at {\it Herschel}/SPIRE
bands and APEX 870 \um compared to the disk SED model. Using the
parameters of the Schechter function in \citet{carniani15}, the
probability of a galaxy with an 1.3 mm flux of $>$0.81 mJy within
4\arcsec\ of the star is $\sim$0.5\%, but increases to 5\% and 14\%
chance within the FWHM and 10\% of the primary beam, respectively. For
the fainter source, the probability of having a source with an 1.3 mm
flux of $>$0.1 mJy within 4\arcsec\ of the star is 11.4\%. Assuming
the two sources are physically not related, the chance of both within
4\arcsec\ of the star is 0.06\%. The contamination from multiple
background point sources is also seen around $\epsilon$ Eri
\citep{chavez16}.  Therefore, it is possible that the sources in HD~95086 
are related background galaxies, and the 0.06\% probability is
then a lower limit. These values are a statistical assessment given an
ensemble of observations, and the application to one single
observation is not one-to-one correspondence. Based on these
probabilities, the faint source is very likely to be a background
galaxy; the bright source could be a background galaxy, a rare case
but not impossible.

\begin{figure}[thb] 
  \figurenum{7}
  \epsscale{1.1}
  \label{fig_spindex}    
  \plotone{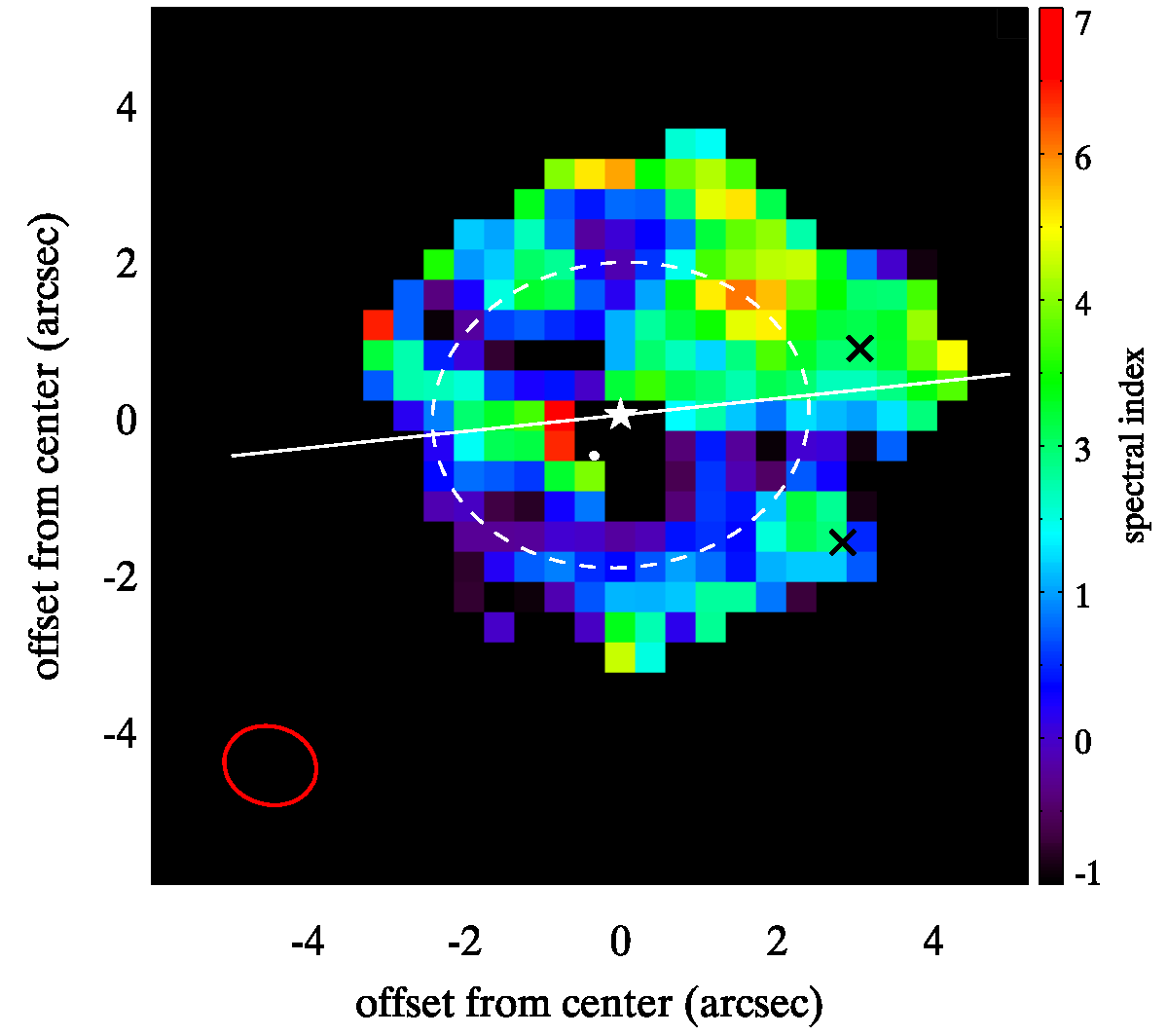}
  \caption{ The spectral index map of the combined data set. To
enhance the signal-to-noise in the spectral index map, a pixel of
0\farcs4 was used. The star, planet b, disk major axis and the two
point sources are all marked as in previous figures. }
\end{figure}

In general, a background galaxy might have a steep dust spectrum, like
$\nu^{3.5}$, and the debris disk is likely to be shallower, like
$\nu^{2.6}$ \citep{macgregor16a,holland17}. Therefore, a background
galaxy is likely to be brighter in the data set B than in A due to the
frequency difference if the absolute flux calibration in both data
sets is consistent. As estimated in Section 4.1, the
flux difference of the bright source between the two data sets is
within the uncertainty of absolute flux calibration, i.e., not
significant. Alternatively, we can compare the spectral indexes
between the bright source and the disk. Spectral index maps were
generated with the data sets combined and separately using all
baselines and long baselines only. Figure \ref{fig_spindex} shows the
spectral index map of the combined data set. The bright source has a
spectral index of 3.0$\sim$4.0 derived from the combined data. Due
to the signal-to-noise in the extended emission, the spectral index
across the whole disk varies, however, the index of the disk is
generally shallower than the one of the bright source.

A steeper spectral index does not necessarily mean the bright source
is indeed a background galaxy because an impact produced clump may
also have a steep particle size distribution, resulting in a high
spectral index. Assuming this is indeed the case, the spectral slope
suggests that the clump should have a total flux of a few hundred mJy
at 200 $\mu$m from extrapolating the measured flux of 0.81 mJy at 1.3
mm. The flux of such a clump at 70--100 $\mu$m range would have been
even brighter, i.e., comparable to the total disk flux in the far
infrared. Given the measured disk SED (Figure \ref{sed}), it would be
very difficult to have such a component co-exist with other components
(planetesimal disk and disk halo), corroborating our early assessment.

\begin{figure}[thb] 
  \figurenum{8} 
  \label{galaxysed} 
  \plotone{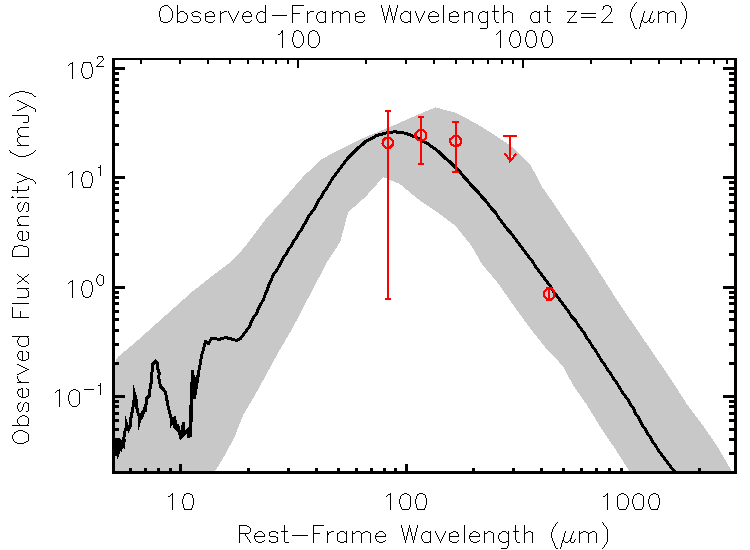}
  \caption{The possible SED of the background galaxy at 250, 350, 500
\um (SPIRE bands) and 1.3 mm. The flux density in the SPIRE bands is
estimated by the excess emission between the observed values and the
SED model, and the 1.3 mm flux density is directly measured from the
ALMA data. The black line shows a representative SED for a submillimeter
galaxy from \citet{pope08}, and the gray area shows the plausible
range of background galaxies of fixed infrared luminosity 10$^{12.5}
L_{\sun}$ (adopted from \citet{casey14}.)}
\end{figure}

As a sanity check, we can also construct the SED of the bright source
using the measured 1.3 mm flux and the revised 3-component disk SED
presented in Section 4.2. By comparing the photometric measurements
and the model disk SED, the ``excess'' emission, presumably from the
bright source, is 31.4$\pm$18.9 mJy, 30.9$\pm$10.6 mJy, 24.8$\pm$10.4
and 10.1$\pm$11.0 mJy at 250, 350, 500 and 870 \mm, respectively (the
quoted errors include 10\% uncertainty from the SED model). With the
1.3 mm flux measured by ALMA, the SED of the bright source is shown in
Figure \ref{galaxysed}, and the SED is consistent with the one of a
dusty star forming galaxy at $z$=2 \citep{casey14}. The angular size
of the bright source (Section 3.4) is in the range of angular sizes
for luminous infrared and submillimeter galaxies at $z\sim$2 measured
in the radio \citep{gurvits99,rujopakarn16}. It seems very plausible
that this bright source is due to the chance alignment of a background
galaxy.

\subsection{Constraints for the HD~95086 b from the Disk Perspective}

HD~95086 b was discovered by \citet{rameau13} at an angular separation
of 0\farcs60--0\farcs63 from the star \citep{rameau16}. At a distance
of 83.8 pc, this translates to a projected distance of 50.3--52.3
au. Assuming the planet b and the disk are co-planar, the inclination
of the disk (30\arcdeg$\pm$3\arcdeg) implies that planet b has a
stellocentric distance of 56--63 au ($\sim$semimajor distance if the
planet b is on a circular orbit).  The bright source makes it difficult
to assess the asymmetry in the disk, and no ``significant'' asymmetry
is present in the millimeter disk emission. Alternatively, we can also
put some constraint on the eccentricity of the shepherding planet orbit,
presumably the planet b, by determining the offset of the ring and the
star (i.e., the offset is $\sim a e$ where $a$ is the semi-major
distance and $e$ is the eccentricity). As discussed earlier, we did
not detect a significant offset between the ring center and the star.
The expected pointing accuracy, $\sim$ resolution/signal-to-noise, is
0\farcs13 since the main ring is detected at S/N$\gtrsim$10. The
non-detection of an offset suggests $e<0.17$ for the shepherding
planet (presumably planet b) with a semi-major distance of 63 au.

The most recent orbital parameters for HD~95086 b are from
\citet{rameau16} where a small angular movement is detected using
the data obtained by Gemini/GPI between 2013 and 2016: a semi-major distnce
of 61.7$^{+20.7}_{-8.4}$ au, an inclination of 27\arcdeg$^{+10}_{-13}$
and an eccentricity less than 0.2. With the revised distance of 83.8
pc to HD~95086, we revise the semi-major distance of HD~95086 b to
57.2$^{+19.2}_{-7.8}$ au.  These orbital parameters are all consistent
with the ones derived from the disk geometry (assumed co-planar).
The mass of the planet is estimated to be 4.4$\pm$0.8 $M_{Jup}$
\citep{derosa16}. The 5-$\sigma$ detection limits from VLT/NaCo and
Gemini/GPI observations suggest that our current high contrast capability
is not sensitive to planet masses less than 1.5 $M_{Jup}$ in the
60--800 au region from the star \citep{rameau16}.

Assuming the mass of the star is 1.6$\pm$0.16 $M_{\sun}$, the mass
ratio between the planet b and the star, $\mu \equiv M_p/M_{\ast}$, is
1.95--3.45$\times10^{-3}$. For this mass ratio, the timescale to clear
the planet's chaotic zone is $\lesssim 1$ Myr \citep{morrison15}, much
shorter than the estimated $\sim$17 Myr age of the system. We adopt
the numerically derived formula from \citet{morrison15} to compute 
the size of the
planet's chaotic zone. Assuming the planet b is at a circular orbit
with a semi-major axis of $a_p$, the interior chaotic zone width is $\Delta
a_{int}=1.2 \mu^{0.28} a_p$ and the exterior chaotic zone width is $\Delta
a_{ext}=1.7 \mu^{0.31} a_p$. With the range of $\mu$ and $a_p$ (56--63
au), the width of the exterior chaotic zone is 14--19 au, suggesting
the outer boundary of the chaotic zone from the planet b is 70--82
au. If the inner edge of the disk is at 106 au as derived from the two
boundary model in Section 3.4, an eccentricity of $e\sim$0.29 is
needed to extend its chaotic zone to the inner edge of the disk if
planet b is the shepherding planet and coplanar with the disk. This
eccentricity is marginally consistent ($<$2$\sigma$) with no significant offset
detected between the ring center and the star. If we relax the
assumption that the disk and the planet are co-planar, HD~95086 b's
chaotic zone can reach 98 au assuming a semi-major distance of 76 au
(the maximum allowable range from high contrast imaging,
\citealt{rameau16}). A small eccentricity ($e\sim$0.08) can extend the
planet's influence to 106 au (the inner boundary of the
disk). Therefore, the planet b can be the shepherding planet to
maintain the inner edge of the Kuiper-belt analog.  In summary, the
values listed above are all within the allowable ranges for the
system.

If the system hosts an additional planet outside the orbit of HD~95086
b that is shepherding the inner edge of the cold disk, it would have
to be less than 1.5 $M_{Jup}$ to have eluded detection
\citep{rameau16}. From the dynamical stability criteria of
\citet{gladman93} and \citet{morrison16}, the estimated separation
between planet b and the disk is sufficient for another planet below
this mass threshold to reside there while remaining long-term
dynamically stable with respect to planet b if both possess low
orbital eccentricities. However, the youth of this system ($\sim$17
Myr) and long dynamical timescales at these large orbital distances
place a lower limit on planets that could have cleared debris from
their chaotic zones over the system's lifetime. From the clearing
timescales estimated in \citet{morrison15}, a putative low
eccentricity, coplanar outer planet orbiting in the region beyond
planet b's orbit and interior to the cold disk, would have to be
$\gtrsim$0.2 $M_{Jup}$ ($\sim$4 Neptune masses) to have cleared debris
from that region over the system's lifetime. In summary, the
shepherding planet between the planet b and the inner edge of the cold
disk would have a mass of 0.2--1.5 $M_{Jup}$ with the assumption of
coplanarity and low eccentricity.

\section{Conclusion}

We obtained an ALMA 1.3 mm image of HD~95086, a young ($\sim$17 Myr)
star hosting a directly imaged planet and a debris disk composed of
dust generated in massive asteroid- and Kuiper-belt analogs. The high
angular resolution (a beam of 1\farcs1) and sensitivity (rms of 7.5
$\mu$Jy beam$^{-1}$) provided by ALMA enable us to resolve the
Kuiper-belt analog for the first time. The sensitive ALMA millimeter
image reveals an inclined ring centered at the star and a bright
source near the edge of the ring along the major axis of the
ring. Our observations also covered the the $^{12}$CO J=2-1 transition
at 230.538 GHz, and no CO emission above 3 $\sigma$ per beam was found
in the pipeline produced CO channel maps in the region of the ring
and the bright source.

To access the properties of the bright source, we also generated the
continuum map using data with baselines longer than 60--80 $k\lambda$
where an extended structure like the planetesimal disk is filtered
out. The long baseline data reveal two sources within 3\farcs5 of the
star (a much fainter source south of the bright one). The FWHM of the
sources is consistent with being a point source although the bright
one is slightly broader (8\%) than the synthesized beam. We determined
the best fit parameters (total fluxes and positions) of the two
sources in the long baseline data using point source fitting in the
visibility domain. The faint source has a 1.3 mm flux of 0.1 mJy (S/N
$\gtrsim$9 in the long baseline map), and is most likely a background
galaxy, similar to other faint sources in the long baseline map. The
bright source is detected at high S/N with a total 1.3 mm flux of 0.81
mJy and located --3\farcs08 E and 0\farcs83 N of the star. We explored
the possible nature of the bright source including (1) a debris phase
of the circumplanetary disk, (2) a dust clump produced by a giant
impact, (3) a dust concentration due to planetesimals trapped by an
unseen planet, and (4) a background dusty galaxy. The source's
brightness in dust continuum and non-detection of the CO emission
suggest that it is unlikely to have resulted from a structure
physically assocated with the system (the first three scenarios), and
more likely due to chance alignment of a background source.  We
further constructed the SED of the bright source using the 1.3 mm flux
and the ``excess'' emission by comparing the unresolved photometry and
the expected disk emission, and found it is consistent with the
expected SED from a $z$=2 dusty galaxy. The slight extension of the
bright source and the steeper spectral index compared to the spectral
index of the disk are both consistent with the bright source being a
luminous, high redshift galaxy.

We used the MCMC approach to determine the best fit parameters of the
disk. We assumed the Kuiper-belt analog can be described by simple,
parametric models.  We explored two axi-symmetric, geometric thin
models for the disk surface density profiles: (1) a two-boundary,
power-law disk with sharp inner and outer boundaries, and (2) a
Gaussian ring. We further assumed that the millimeter emission comes
from large grains whose temperatures follow a $r^{-0.5}$ power
law. The best-fit parameters and associated uncertainties are derived
by fitting the visibilities and image plane data. We found that
fittings in the visibilities and image plane domain give consistent
results within $\pm$1 $\sigma$ uncertainties in terms of the geometric
parameters (disk extent, position and inclination angles and the
offset of the bright source); however, the derived fluxes are
consistently larger when using the fits in the image plane. Although
within the uncertainties, the flux of the bright source is also
brighter than the one derived from the long baseline data, implying a
difficulty of separating the disk from the bright source. We then
synthesized the disk parameters by (1) fixing the point source
parameters in the visibilities fit, (2) masking out the pixels
affected by the bright source in the imaging fit, and (3) fitting the
bright source as a Gaussian profile. The final parameters for the
Kuiper-belt analog are all very similar. The major axis of the disk is
along P.A. of 97\arcdeg$\pm$3\arcdeg\ with the mid-plane inclined by
30\arcdeg$\pm$3\arcdeg\ from face-on. The width of the disk is very
broad and resolved by $\lesssim$2 beam widths. The disk density
profile is consistent with either (1) a broad, Gaussian ring peaked at
200$\pm$6 au with a FWHM width of 168$\pm$7 au or (2) an
$r^{-0.5\pm0.3}$ power-law profile with an inner radius of 106$\pm$5
au and outer radius of 320$\pm$10 au. Although the residual maps (data
-- model) are very similar between the two models, the Gaussian ring
model gives a slightly better reduced $\chi^2$. In all residual maps,
the east side of the disk has more negative residuals compared to the
opposite side, suggesting an apparent disk asymmetry. However, the
residual in the west side of the disk is also contaminated by the
imperfect subtraction of the bright source. Judging from the azimuthal
profile along the disk circumference, the apparent asymmetry is not
significant.

We also explored whether allowing an offset between the ring and the
star would produce a better residual map. Although a small offset
(within 0\farcs15, $\sim$one eighth of the synthesized beam) is
preferred in $\chi^2$ statistics, it produces no significant
difference in the residual map. Based on the estimated pointing
accuracy of 0\farcs13, the non-detection of an offset suggests that
the orbit of the shepherding planet has an eccentricity $<$0.17 if it has a
semi-major axis of 63 au. Given the observed projected separation
between HD~95086 b and the star, the semi-major axis of the planet b orbit
is 56--63 au if the planet and the disk share the same orbital plane
and the planet is on a circular orbit. The estimated major axis is
consistent with the apparent motion of the planet detected within
three years. However, in the co-planar case for a planet in a circular
orbit, the expected chaotic zone of the planet b (maximum of 82 au),
does not reach the inner boundary of the disk ($\sim$106 au). If the
planet b is the shepherding planet to maintain the inner edge of the
cold disk, an eccentricity $\gtrsim$0.29 is needed to extend its
influence. Such a planet would create an offset of 0\farcs2 between
the star and the ring center, marginally consistent with the observed
data. It is also possible that the planet b is not the shepherding
planet for the cold disk given the large separation between its
chaotic zone and the inner disk edge. An additional unseen low-mass
planet (0.2--1.5 $M_{Jup}$) on a circular orbit can also occupy the
dust-free zone between the planet b and the cold disk and have eluded
detection. Alternatively, relaxing the co-planarity assumption, a
larger semimajor axis of planet b (76.4 au, still within the allowable
range) and a small eccentricity ($\sim$0.08) would extend its chaotic
zone to $\sim$106 au.

\acknowledgments

We thank Alan Jackson for the discussion on the evolution of impact
produced clumps, and Benjamin Weiner for the discussion on the
properties of background galaxies. We are grateful to George Rieke for
his thorough proofreading of the manuscript. We also thank Julien
Rameau providing additional inforamtion about the background source,
and Scott Kenyon and Ya-Lin Wu for their comments. M.B. is grateful to
Bruce Sibthorpe and Andr\'es Jord\'an for the help during proposal
preparation. K.Y.L.S. acknowledges the partial support from the NASA
grant \# NNX15AI86G.  M.A.M. acknowledges support from the National
Science Foundation under Award No.\ 1701406. M.B. acknowledges support
from the Deutsche Forschungsgemeinschaft (DFG) through project Kr
2164/15-1. R.M. acknowledges research support from NASA (grant \#
NNX14AG93G) and NSF (grant \#AST-1312498).

This paper makes use of the following ALMA data: ADS/JAO.ALMA
\#2013.1.00773.S and \#2013.1.00612.S. ALMA is a partnership of ESO
(representing its member states), NSF (USA) and NINS (Japan), together
with NRC (Canada) and NSC and ASIAA (Taiwan) and KASI (Republic of
Korea), in cooperation with the Republic of Chile. The Joint ALMA
Observatory is operated by ESO, AUI/NRAO and NAOJ. The National Radio
Astronomy Observatory is a facility of the National Science Foundation
operated under cooperative agreement by Associated Universities, Inc.

This work has made use of data from the European Space Agency (ESA)
mission {\it Gaia} (\url{https://www.cosmos.esa.int/gaia}), processed by
the {\it Gaia} Data Processing and Analysis Consortium (DPAC,
\url{https://www.cosmos.esa.int/web/gaia/dpac/consortium}). Funding
for the DPAC has been provided by national institutions, in particular
the institutions participating in the {\it Gaia} Multilateral Agreement.

\medskip 

\noindent Facilities: \facility{Atacama Large Millimeter/submillimeter Array (ALMA)}.

\end{document}